\documentclass[reqno]{amsart}
\usepackage{graphicx,graphics}
\usepackage{subfig}

\theoremstyle{definition}

\numberwithin{thm}{section}
\numberwithin{lem}{section}
\numberwithin{coll}{section}
\numberwithin{rem}{section}
\numberwithin{exm}{section}
\numberwithin{prop}{section}
\numberwithin{equation}{section}
\numberwithin{equation}{section}

\begin{document}

\centerline {\textsc {\large Galton-Watson Process for a class of distributions}}
\centerline {\textsc {\large from Bernoulli to Poisson}}
\vspace{0.5in}

\begin{center}
   R. Vasudeva and 
   Ali Saeb \footnote{Corresponding author: ali.saeb@gmail.com}\\
   Department of Studies in Statistics\\
University of Mysore\\
Mysore 570006, India
\end{center}

\vspace{1in}


\noindent {\bf Abstract:}	In this paper the Galton Watson branching process has been studied for a class of offspring distributions which are in a way sandwiched between the Bernoulli and Poisson.
\vspace{0.5in}

\vspace{0.2in} \noindent {\bf Keywords:} Branching Process, Galton Watson Process, Poisson model, Janardan model, Bernoulli model.

\vspace{0.5in}

\vspace{0.2in} \noindent {\bf AMS subject classification:} 60J80



\newpage
\section{Introduction}
	A branching process is a stochastic model used for the study of propagation of species over generations. One of the simplest form of  branching processes is the Galton-Watson process.
In this process, the following underlying assumptions are made. Any individual of the species, can reproduce and the number of such individuals produced will be a random variable (rv), known as the offspring rv. The number of offsprings produced by two or more members will be mutually independent identically distributed rvs, irrespective of whether the members are of the same generation or of different generations.

Let $\xi$ denote the offspring rv with the probability generating function (pgf) $P(s)$ $=\sum_{i=0}^{\infty}{p_i s^{i}},\;\; 0\leq s \leq 1,$ where $p_i=\Pr(\xi=i),\;\;i\geq 0.$ Denote by $X_n$, the size of the $n^{th}$ generation. It is well known that $(X_n)$ is a Markov chain. Let the pgf of $X_n$ be denoted by $P_n(s).$ Then the relation, $P_n(s)=P_{n-1}(P(s)),$ $0\leq s\leq 1,$ is also well known. The members of $X_0$ are called ancestors. With no loss generality $X_0$ is taken as 1. Some of the characteristics of study are the size of the $n^{th}$ generation, eventual extinction or explosion of the species and so on. In the literature, it has been shown that these characteristic events depend on the offspring distribution. 

A Galton-Watson branching process is said to be sub-critical, critical or super-critical according as $E \xi <1,\;$ $=1\;$ or$\; >1.$ The probability of extinction is 1 if $E\xi<1$ and it is the fractional root of the equation $P(s)=s,\;$ if $E\xi>1,$ (see for example, Karlin and Taylor (1981)). We hence note that most of the characteristics of study depend on the offspring distribution. Ramiga (1977) investigated the behaviour for offspring distributions, such as Poisson, Binomial, Geometric, Pearson family and so on. In this article, we study the Galton-Watson process, assuming the two parameter distribution obtained in Janardan (1980) as the offspring distribution (see, also Janardan et al. (1995)).

The speciality of this class of distributions is that it gives a family of distributions sandwiched between Bernoulli and Poisson. We call these distributions as Perturbed Poisson distributions.

\section{Perturbed Poisson Distribution}
Janardan (1980) in his paper on the oviposition tactics of weevils on beans, obtained a two parameter distribution with probability mass function (pmf)
	\begin{eqnarray} \label{e1}
		p_0&=&e^{-\lambda},\nonumber\\
		p_1&=&\dfrac{\lambda}{\mu-\lambda}(e^{-\lambda}-e^{-\mu}),\nonumber\\
		\vdots &&\nonumber\\
		p_m&=&\dfrac{\lambda \mu^{m-1}}	{(\mu-\lambda)^m}\left(e^{-\lambda}-e^{-\mu}\sum_{j=0}^{m-1}\dfrac{(\mu-\lambda)^j}{j!}\right),\nonumber
	\end{eqnarray}
	$m\geq 2,\;\; \lambda>0,\;\; 0<\mu<\lambda$ with the pgf
	\begin{equation} \label{e2}
		P(s;\mu,\lambda)=\dfrac{1}{\mu(s-1)+\lambda}\left((s-1)(\mu-\lambda)e^{-\lambda}+s\lambda e^{-\mu(1-s)}\right),\;\; 0\leq s\leq 1.
	\end{equation}
	Janardan (1980), in the study of oviposition of beetles on mange beans, observed that the beetles are selective in laying eggs, in the sense that the chance of laying the second and subsequent eggs on a mange bean already having one egg is smaller than the chance of laying an egg on a bean with no egg. Accordingly, he introduced a parameter $\mu$ which can cut down the chance by any desirable extent. It is interesting to note that $\lim_{\mu\rightarrow\lambda} P(s;\mu,\lambda)=e^{-\lambda(1-s)}$ and $\lim_{\mu\rightarrow 0}P(s;\mu,\lambda)=e^{-\lambda}+s(1-e^{-\lambda}),$ which are respectively the pgf of Poisson (with parameter $\lambda$) and Bernoulli (with parameter $(1-e^{-\lambda})$) respectively. 
	
There are micro organisms which give birth to one offspring and die. If the atmosphere is not congenial for its survival, it may die before giving birth to an offspring. In such a case, a Bernoulli model is appropriate for the offspring distribution. Many workers have used Poisson distribution as offspring distribution and studied the Branching process. The two parameters distribution considered in this paper may possibly give models which are more appropriate then the Poisson model as they are, in some sense, sandwiched between the Bernoulli and Poisson. With this motivation, we are studying Galton-Watson model, assuming that the offspring rv, follows a Perturbed Poisson distribution. 
	
 We have,
	\begin{eqnarray}
			E\xi&=&{\dfrac{d}{ds}P(s;\mu,\lambda)\mid_{s=1}},\nonumber\\
			&=&\dfrac{\mu}{\lambda}\{e^{-\lambda}+\lambda-1\}+(1-e^{-\lambda}),\;\; 0<\mu<\lambda.\label{e2.5}
	\end{eqnarray}
	If may be trivially noted that $E\xi$ is an increasing function of $\mu$ with
\[\lim_{\mu\to 0}E\xi=(1-e^{-\lambda})\qquad \text{and} \qquad \lim_{\mu \to\lambda}E\xi=\lambda.\]
	which respectively are the expected values of the Bernoulli and Poisson distributions obtained in limit. Note that, $E\xi<1$ whenever $\;\lambda<1.\;$ Consequently, the branching process is subcritical whenever $\lambda<1.$ In turn the eventual extinction has probability $1.$ When, $\;\lambda\geq 1,\;E\xi<1$ if $\mu<\dfrac{\lambda e^{-\lambda}}{e^{-\lambda}-(1-\lambda)};$ $E\xi=1$ if $\mu=\dfrac{\lambda e^{-\lambda}}{e^{-\lambda}-(1-\lambda)},$ and $E\xi>1$ if $\mu> \dfrac{\lambda e^{-\lambda}}{e^{-\lambda}-(1-\lambda)}.$ The branching process is subcritical, critical or supercritical according as $\mu<g(\lambda),$ $\mu=g(\lambda)$ and $\mu>g(\lambda)$ where $g(\lambda)=\dfrac{\lambda e^{-\lambda}}{e^{-\lambda}-(1-\lambda)}.$
	The probability of extinction is one when $\mu< g(\lambda)$ and is the fractional root of $P(s;\mu,\lambda)=s$ when $\mu>g(\lambda),$ where,
$P(s;\mu,\lambda)$ $=\dfrac{1}{\mu(s-1)+\lambda}\left((s-1)(\mu-\lambda)e^{-\lambda}+s\lambda e^{-\mu(1-s)}\right).$

	The probability of extinction for different values of $\lambda$ and $\mu$ have been obtained by fixed point iteration technique (see, for example, Sastry (2000)) and a comparison has been made with Poisson as offspring model. It may be noted, when the offspring distribution is Bernoulli, the system always extincts. Consequently, even when $\lambda>1$, if $\mu$ is close to zero, extinction occurs with probability $1.$
	
	When the branching process is sub-critical, another interesting measure is the time of extinction. Let $T$ denote the time of extinction. Since, extinction occurs with probability $1,$ $T$ is a proper rv. Recalling that $X_n$ is the size of $n^{th}$ generation, we have $\Pr(T=n)=\Pr(X_n=0,X_{n-1}>0)$. Note that
\begin{eqnarray}
		\Pr(X_n=0)&=&\Pr(X_n=0,X_{n-1}=0)+\Pr(X_n=0,X_{n-1}>0),\nonumber\\
		&=&\Pr(X_{n-1}=0)+\Pr(X_n=0,X_{n-1}>0).\nonumber
\end{eqnarray}
	Hence,
\begin{eqnarray}
		\Pr(T=n)&=&\Pr(X_n=0)-\Pr(X_{n-1}=0),\nonumber\\
		&=&q_n-q_{n-1},\label{e4}
\end{eqnarray}
where $q_n=\Pr(X_n=0)=P_{n-1}(P(0)),\;\;n\geq 2,$ $q_1=\Pr(\xi=0)=P(0),$ and $q_0=\Pr(X_0=0)=0.$
	
\section{Estimation of Parameters}
	For fitting the model to any branching process, the parameters have to be estimated. If $f_0,\,f_1,\ldots$ are the class frequencies based on a random sample of size $n,$ the likelihood function will be given by
	$L(\lambda,\mu)=\Pi_{i=0}^{\infty}p_i^{f_i}$ or
\begin{eqnarray}
\log L(\lambda,\mu)&=&-f_0\lambda+f_1\log\left(\frac{\lambda}{\mu-\lambda}(e^{-\lambda}-e^{-\mu})\right)\nonumber\\
&&+\sum_{m=2}^{\infty}f_m\log\left(\dfrac{\lambda\mu^{m-1}}{(\mu-\lambda)^m}\left(e^{-\lambda}-e^{-\mu}\sum_{j=0}^{m-1}\frac{(\mu-\lambda)^j}{j!}\right)\right).\label{e4.5}
\end{eqnarray}
One can see that the likelihood equations are $\frac{\partial\log L(\mu,\lambda)}{\partial\lambda}=0$ and $\frac{\partial\log L(\mu,\lambda)}{\partial\mu}=0$

where,
\begin{eqnarray}
\frac{\partial\log L(\mu,\lambda)}{\partial\lambda}&=&-f_0+f_1\left(\frac{1}{\lambda}+\frac{1}{\mu-\lambda}-\frac{e^{-\lambda}}{e^{-\lambda}-e^{-\mu}}\right)\nonumber\\
&&+\sum_{m=2}^{\infty}f_m\left(\frac{1}{\lambda}+\frac{m}{\mu-\lambda}-\frac{e^{-\lambda}-e^{-\mu}\sum_{j=2}^{m-1}(\mu-\lambda)^{j-1}/(j-1)!}{e^{-\lambda}-e^{-\mu}\sum_{j=0}^{m-1}(\mu-\lambda)^j/j!}\right),\nonumber\\
&=&-f_0+\frac{\sum_{m=1}^{\infty}f_m}{\lambda}+\frac{\sum_{m=1}^{\infty}mf_m}{\mu-\lambda}-\frac{e^{-\lambda}f_1}{e^{-\lambda}-e^{-\mu}}\nonumber\\
&&-\sum_{m=2}^{\infty}\frac{f_me^{-\mu}}{e^{-\lambda}-e^{-\mu}\sum_{j=0}^{m-1}(\mu-\lambda)^j/j!}.\label{e5}
\end{eqnarray}
and,
\begin{eqnarray}
\frac{\partial\log L(\mu,\lambda)}{\partial\mu}&=&f_1\left(\frac{e^{-\mu}}{e^{-\lambda}-e^{-\mu}}-\frac{1}{\mu-\lambda}\right)\nonumber\\
&&+\sum_{m=2}^{\infty}f_m\left(\frac{m-1}{\mu}-\frac{m}{\mu-\lambda}+\frac{\frac{(\mu-\lambda)^{m-1}e^{-\mu}}{(m-1)!}}{e^{-\lambda}-e^{-\mu}\sum_{j=0}^{m-1}\frac{(\mu-\lambda)^{j}}{j!}}\right).\label{e5.5}
\end{eqnarray}

From the expressions, one can see that solving for $\lambda$ and $\mu$ is highly complicated. Coming to moment estimation, $E\xi$ and $Var(\xi)$ are to be equated to the sample mean and sample variance. We have (see, Janardan (1980)) 
\[Var(\xi)=\mu^2-E^2\xi+(1-e^{-\lambda})\left(1-\frac{\mu}{\lambda}\right)\left(1-\frac{2\mu}{\lambda}\right)+\mu\left(3-\frac{2\mu}{\lambda}\right).\]
The expression for $Var(\xi)$ is very involved, thus ruling out the moment estimation technique.

We hence adopt the following method of repeated moment estimation. We know that the offspring rv $\xi$ takes values $0,1,2,\ldots$ Define a new rv $Z$ with
	\begin{eqnarray*}
Z= \left\lbrace	
	\begin{array}{l l}
	0, &\;\text{if}\;\; \xi\geq 1,\\
	1, &\;\text{if}\;\; \xi=0. \\
 \end{array}
  \right.					
\end{eqnarray*}
i.e. $Z=1$ if there are no offspring and $=0$ if there are one or more offspring. $Z$ is a Bernoulli rv with $\Pr(Z=1)=e^{-\lambda}$ and $\Pr(Z=0)=1-e^{-\lambda}.$ Thus $EZ=e^{-\lambda}.$ Hence $\lambda$ is estimated by moment estimation, i.e. from the equation, $EZ=\bar{Z},$ where $\bar{Z}=\frac{\sum_{j=1}^{\infty}Z_j}{n}=\frac{f_0}{n}.$ The solution is given by 
\begin{eqnarray}
\hat{\lambda}=\log n-\log f_0.\label{e6}
\end{eqnarray}
Having obtained $\hat{\lambda},$ $\hat{\mu}$ is obtained by equating the first moment $E\xi$ with the sample mean $\bar{X}=\frac{\sum_{j=1}^{\infty}jf_j}{n}$ and substituting $\hat{\lambda}$ for $\lambda.$ We get
\begin{eqnarray}
\hat{\mu}=\frac{\hat{\lambda}(\bar{X}-1+e^{-\hat{\lambda}})}{e^{-\hat{\lambda}}+\hat{\lambda}-1}.\label{e7}
\end{eqnarray}
Such a technique has been suggested in Anscombe (1950) and the same is also used in Janardan (1980). In what follows, we introduce the algorithm for estimating the parameters. 

Algorithm:
\begin{enumerate}
\item Generate $n$ observations from Janardan distribution.
\item Count the number of $0$ from the data generated and call it as $f_0$.
\item Compute the value of mean from the data and call it as $\bar{x}.$
\item Estimate $\lambda$ and $\mu$ from (\ref{e6}) and (\ref{e7}).
\end{enumerate}
We introduce the R program in Appendix A and B for generating the observations and estimating $\lambda$ and $\mu.$  $1000$ observations each were generated from populations with (i) $\lambda=0.8$ and $\mu=0.4,$ (ii) $\lambda=2,$ $\mu=1.9$ and (iii) $\lambda=2,$ $\mu=1$ and $f_0$ and $\bar{x}$ were computed in each case. By equations (\ref{e6}) and (\ref{e7}), the estimates of $\lambda$ and $\mu$ were obtained. Table \ref{T5} gives the parameters and the estimates.

Assuming that the data with $\lambda=2,$ $\mu=1.9$ is from Poisson ($\lambda$), the estimate $\hat{\lambda}$ is obtained as $\hat{\lambda}=1.948.$

\section{Comparison with Poisson model}
	The probability of extinction for Poisson model (PM) and Janardan model (JM) for different values of $\lambda$ and $\mu$ are given in Table \ref{T1}. These values are obtained by solving the equations $e^{-\lambda(1-s)}=s$ for PM and $P(s;\mu,\lambda)=s$ for JM. It may be seen that the probability of extinction in Poisson model goes to zero when the value of $\lambda$ increases. Also, in Janardan model when the values of $\mu$ approach $\lambda,$ then the probability of extinction would be close to that of Poisson model.

	Table \ref{T2} gives the probability of extinction under Poisson model for $20$ generations. We observe that the probability of extinction for $\lambda=0.8,$ goes to $1,$ and for $\lambda=2,$ it slowly increasing and stabilizes at $0.2032$ after $18th$ generation. The probability of explosion is nearly $0.8.$ For $\lambda=8,$ the probability of extinction is negligible and the probability stabilizes after $4th$ generation. Here, the species size eventually explodes with probability $0.99.$

	In Table \ref{T3}, we have the probability of extinction for Janardan model for $20$ generations. We observe that, the probability of extinction for $\lambda=0.8$ and $\mu=0.4$ goes to $1.$ When $\lambda=2,$ we compute the value of $\mu_0=\dfrac{\lambda e^{-\lambda}}{e^{-\lambda}-(1-\lambda)}.$ One can see that, the model is sub-critical for $\mu<\mu_0,$ and super critical for $\mu>\mu_0.$ The probability of extinction for $\mu=0.2$ is $1$ and for $\mu>0.2384$ it is less than $1.$ For the $20th$ generation, the size will be $0$ with probability $0.9997$ when $\lambda=0.8,$ $\mu=0.4;$ it is $0.3060074$ for $\lambda=2$ $\mu=1$ and $0.2083325$ for $\lambda=2$ $\mu=1.9$. Another interesting problem is that of finding the probability of extinction in $n\,th$ generation, as $n$ grows.

Table \ref{T4} gives the values of the extinction probabilities for generations upto $20.$ In (\ref{e4}), $T$ denotes a proper rv and extinction occurs with probability $1.$ We study the model when it is sub-critical. The Poisson model is sub-critical where $\lambda<1$ and Janardan model is sub-critical when $\lambda<1,$ or $\lambda>1,$ $\mu<\mu_0.$
	Figure 1 shows the probability of extinction for JM with $\lambda=2$ and with the values of $\mu=0.2,1$ and $1.9$ and for Poisson model with $\lambda=2.$

\clearpage
\appendix
\section{Tables and Graph}
	\begin{table}[!ht]
		\centering
		\caption{Parameters and their estimates for Three sets of data generated from JM}\label{T5}
		\begin{tabular}{|c|c|c|c| }
		\hline
		$\lambda$ &$\mu$& $\hat{\lambda}$ &$\hat{\mu}$ \\
		\hline
		$0.8$&$0.4$&$0.79186$&$0.44628$\\
		$2$&$1.9$&$1.95193$&$1.93778$\\
		$2$&$1$&$1.9951$&$0.9948$\\
		\hline
  	\end{tabular}
  \end{table}

\begin{table}[!ht]
		\centering
		\caption{The probability of extinction for Poisson and JM}\label{T1}
		\begin{tabular}{|l|l|l|}
			\hline
			$\lambda$  & PM & JM $(\mu)$ \\
			\hline
			$1.5$  & $0.4172135$  & $0.4172301\,(1.4999)$ \\
			$2$  & $0.2032028$  & $0.2032077\,(1.9999)$ \\
			$3$  & $0.05952168$  & $ 0.05952225\,(2.9999)$ \\
			$4.5$  & $0.01171188$  & $0.01171191\,(4.49999)$ \\
			$6$ &  $0.002517337$  & $0.002517339\,(5.9999)$ \\
			\hline
		\end{tabular}\end{table}
	
	\begin{table}[!ht]
			\centering
			\caption{Probability of extinction for $20$ generation in Poisson model}\label{T2}
			\begin{tabular}{|c|c|c|c|}
			\hline
			Generation & $\lambda=0.8$ & $\lambda=2$ & $\lambda=8$ \\
			\hline
$1$&$0.4493289641$ & $0.1353352832$ &$0.0003354626$\\
$4$& $0.8200277487$& $0.1990798058$ & $0.0003363666$\\
$5$& $0.8659069701$&$ 0.2015252917$&$ 0.0003363666$\\
$10$& $0.9628438404$& $0.2031694953$& $0.0003363666$\\
$15$& $0.9884060102$ &$0.2031876663$ &$0.0003363666$\\
$20$& $0.9962592647$ &$0.2031878677$ &$0.0003363666$\\
			\hline
		\end{tabular}
	\end{table}

	\begin{table}[!ht]
		\centering
		\caption{Cumulative Probability of extinction for $20$ generations in JM}\label{T3}
		\begin{tabular}{|c|c|c c c| }
		\hline
		Generation &  $\lambda=0.8$ & & $\lambda=2$ &  \\
& $\mu=0.4$ & $\mu=0.2$ & $\mu=1$ & $\mu=1.9$ \\
		\hline
		$1$ & $0.4493290$ & $0.1353353$ & $0.1353353$ & $0.1353353$ \\
	    $5$ & $0.9112714$ & $0.4477756$ & $0.2796819$ & $0.2061979$ \\
	    $10$ & $0.9881263$ & $0.6303605$ & $0.3028484$ & $0.2083028$ \\
	    $15$ & $0.9983430$ & $0.7296663$ & $0.3056545$ & $0.2083321$ \\
	    $20$ & $0.9997675$ & $0.7919229$ & $0.3060074$ & $0.2083325$ \\
		\hline
		\end{tabular}
	\end{table}

	\begin{table}[!ht]
		\centering
		\caption{Probability of extinction upto $20$ generations in PM and JM}\label{T4}
		\begin{tabular}{|c|l l|l l l| }
		\hline
		Generation &PM& $\lambda=0.8$ & JM &$\lambda=0.8$($\mu=0.4$) &$\lambda=2$($\mu=0.2$)\\
		\hline
		$1$&&$4.49e-01$&&$4.49e-01$&$0.135335283$\\
		$5$&&$4.59e-02$&&$4.64e-02$&$0.056564229$\\
		$10$&&$1.02e-02$&&$5.77e-03$&$0.027352681$\\
		$15$&&$2.98e-03$&&$7.98e-04$&$0.01605384$\\
		$20$&&$9.44e-04$&&$1.12e-04$&$0.01050924$\\
		\hline
  	\end{tabular}
  \end{table}

\begin{figure}[!ht]
  \centering
   \subfloat{\includegraphics[width=1\textwidth]{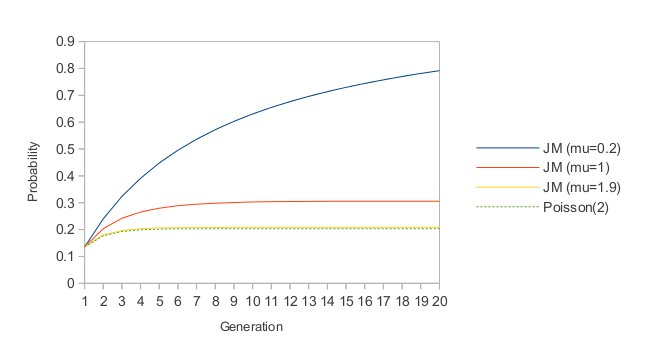}} \label{Graph}
    \caption{\small{Probabilities of extinction for JM with $\lambda=2$  and with values of $\mu=0.2,1$ and $1.9$ (lines) and Poisson model with $\lambda=2$ (dash line) for $20$ generations.}}
\end{figure}
\clearpage
\section{} R program to compute the values of tables
\begin{verbatim}
library(rootSolve)
#------ Table 1 (Poisson Model)
for(la in c(1.5,2,3,4.5,6)){f1<-function(s1) exp(-la*(1-s1))-s1 
pm<-uniroot.all(f1,c(0,1)); print(pm)
#------ Table 1 (Janardan Model)
mu<-la-0.0001
f2<-function(s2) ((s2-1)*(mu-la)*exp(-la)
+s2*la*exp(-mu*(1-s2)))/(mu*s2-mu+la)-s2
jm<-uniroot.all(f2,c(0,1)); print(jm)}
#------ Table 2 (Janardan Model)
f1<-function(s1) {exp(-la*(1-s1))}
la<-c(.8,2,8); s1<-exp(-la);print(s1)
for(i in 1:20){s1<-f1(s1); print(s1)}
#------ Table 3 (Janardan Model)
f2<-function(s2) {((s2-1)*(mu-la)*exp(-la)
+s2*la*exp(-mu*(1-s2)))/(mu*s2-mu+la)}
la<-2; mu<-c(.2,1,1.9); s2<-exp(-la); print(s2)
for(i in 1:20){s2<-f2(s2);print(s2)}
\end{verbatim}

\section{}
R program to generate data from Janardan model and estimate the parameters
\begin{verbatim}
la<-5;mu<-.5
g<-array(0)
p<-array(0);nn<-1000
u<-runif(nn);x<-array(0)
g[1]<-p[1]<-exp(-la)
p[2]<-la*(exp(-la)-exp(-mu))/(mu-la)
g[2]<-p[2]+p[1];i<-2
while(round(g[i],5)<1){s<-0
for(j in 0:(i-1))
s<-s+(mu-la)^j/factorial(j)
p[i+1]<-(la*mu^(i-1))/((mu-la)^i)*(exp(-la)-exp(-mu)*s)
g[i+1]<-round(p[i+1],5)+round(g[i],5)
i<-i+1}
n<-i
for(i in 1:nn){for(j in 1:n)
if (u[i]<=g[j]) {x[i]<-(j-1); break}}
f<-array(0,n);j<-1
while(j<=n){s<-0
for(i in 1:nn)
if (x[i]==(j-1)) s<-s+1
f[j]<-s;j<-j+1}
w<-c(0:(n-1));m<-mean(x)
la1<-log(nn)-log(f[1])
mu1<-(la1*(m-1+exp(-la1)))/(exp(-la1)+la1-1)
mu1;la1

#--- Generate data from Poisson and estimate parameters with JM
x2<-rpois(nn,3)
f2<-array(0);j<-1
while(j<=max(x2)){s<-0
for(i in 1:nn)
if (x2[i]==(j-1)) s<-s+1; f2[j]<-s
j<-j+1}
x2_max<-max(x2)
m2<-mean(x2)
la2<-log(nn)-log(f2[1])
mu2<-(la2*(m2-1+exp(-la2)))/(exp(-la2)+la2-1)
mu2;la2
\end{verbatim}
	

\end{document}